\documentclass[prb,showpacs,twocolumn,amsmath,amssymb,superscriptaddress]{revtex4}
\usepackage{epsfig}
\usepackage{graphicx}
\usepackage{dcolumn}
\usepackage{bm}

\begin{document}
\title{Interplay of electron-phonon interaction and strong correlations: 
DMFT+$\Sigma$ study}
\author{E.Z.~Kuchinskii}
\affiliation{Institute of Electrophysics, Russian Academy of
Sciences, Ural Division, 620016 Ekaterinburg, Russia}
\author{I.A.~Nekrasov}
\affiliation{Institute of Electrophysics, Russian Academy of
Sciences, Ural Division, 620016 Ekaterinburg, Russia}
\author{M.V.~Sadovskii}
\affiliation{Institute of Electrophysics, Russian Academy of
Sciences, Ural Division, 620016 Ekaterinburg, Russia}

\date{\today}

\begin{abstract} 

We perform investigation of Hubbard model with interaction between strongly 
correlated conducting electrons on a lattice with Debye phonons.
To solve the problem generalized dynamical mean-field DMFT+$\Sigma$ method is 
employed with ``external'' self-energy $\Sigma_{ph}$ corresponding to 
electron-phonon interaction. We present DMFT+$\Sigma_{ph}$ results for densities 
of states and kinks in  energy dispersions for 
a variety of model parameters, analyzing the interplay of recently discovered 
kinks of purely electronic nature and usual phonon kinks in the 
electronic spectrum.

\end{abstract}


\maketitle 
\section{Introduction}

The problem of the interplay of strong electronic correlations with electron--phonon
interaction is of central importance in the physics of highly
correlated systems. Actually there is rather long history of such studies, e.g.
one of the most popular models for electron-phonon interaction (EPI)
in strongly correlated systems is the so-called Hubbard-Holstein model (HHM).
The Hubbard model\cite{Hubbard} itself describes local Coulomb interaction of 
electrons on a lattice including e.g. Mott-Hubbard metal-insulator transition.
On the other hand Holstein model contains local linear displacement-to-density
interaction of conducting electrons with local (Einstein) phonon modes\cite{Holstein59}.

Active investigations of the properties of the HHM were undertaken in the framework
of dynamical mean-field theory (DMFT)\cite{DMFT_method}, which is non-perturbative approach
with respect to interaction parameters of the Hubbard model.
Among many others one should mention DMFT solution of HHM for the case where impurity
solver used was the numerical renormalization group (NRG)
(see for review of DMFT(NRG) applications Ref.~\onlinecite{NRGrev}).
The mapping of HHM to Anderson-Holstein impurity was first
performed by Hewson and Mayer\cite{HewMay02}. It was shown that using NRG one can
compute in a numerically exact manner total electron-phonon contribution
to the self-energy of the problem, thus making solution of the HHM
non-perturbative also with respect to electron-phonon coupling strength.
One should note that self-consistent set of DMFT equations is preserved in 
this approach.

However, up to now there are apparently no studies of strongly correlated electrons
interacting with Debye phonons. It is even more surprising in view of the widely
discussed physics of kinks in electronic dispersion observed in ARPES experiments
40-70 meV below the Fermi level of high-temperature superconductors\cite{Lanz}, which are
often attributed to EPI \cite{Shen}. To our knowledge problem of kink formation on electronic 
dispersion caused by EPI in strongly correlated systems was briefly discussed 
within HHM in papers by Hague\cite{Hague} and Koller~{\it et al.}\cite{Koller05}.

In this paper we report DMFT+$\Sigma$ results for the Hubbard model supplemented
with Debye phonons, assumung the validity of Migdal theorem (adiabatic
approximation). 
We consider the influence of Debye phonons on the weakly and
strongly correlated electrons, studying electron dispersion and density of
states (DOS), in particular close to Mott-Hubbard metal insulator transition.
We analyze in details how EPI affects electronic dispersions in correlated
metal and discuss the interplay of recently discovered kinks of purely
electronic nature in electronic dispersion \cite{Nature} and usual phonon 
kinks in the electronic spectra.

The paper is organized as follows. First we introduce in Sec.~\ref{comp} 
DMFT+$\Sigma$ approach to the model at hand. Then in Sec.~\ref{results} 
calculated results are presented and discussed. Suumary and conclusions are
given in Sec.~\ref{conclusion}.

\section{
DMFT+$\Sigma$ computational details}
\label{comp}

The major assumption of our DMFT+$\Sigma$ approach is that the lattice
and time Fourier transform of the single-particle Green function 
can be written as:
\begin{equation}
G_{\bf p}(\varepsilon)=\frac{1}{\varepsilon+\mu-\varepsilon({\bf p})-\Sigma(\varepsilon)
-\Sigma_{\bf p}(\varepsilon)}
\label{Gk}
\end{equation}
where $\varepsilon({\bf p})$ is the bare electron dispersion,
$\Sigma(\varepsilon)$ is the {\em local} self--energy of DMFT, 
while $\Sigma_{\bf p}(\varepsilon)$ is some 
``external'' (in general case momentum dependent) self--energy. 
Advantage of our generalized approach is the additive form of 
the self-energy (neglect of interference) in Eq. (\ref{Gk}) 
\cite{fsdistr,dmftsk,opt}. 
It allows one to keep the set of self-consistent equations of standart DMFT 
\cite{DMFT_method}. However there are two distinctions. 
First, on each DMFT iterations we recalculate corresponding  ``external'' 
self-energy $\Sigma_{\bf p}(\mu,\varepsilon,[\Sigma(\varepsilon)])$ within some 
(approximate) scheme, taking into account interactions e.g. with collective 
modes (phonons, magnons etc.) or some order parameter fluctuations.
Second, the local Green's function of effective impurity problem is defined as
\begin{equation}
G_{ii}(\varepsilon)=\frac{1}{N}\sum_{\bf p}\frac{1}{\varepsilon+\mu
-\varepsilon({\bf p})-\Sigma(\varepsilon)-\Sigma_{\bf p}(\varepsilon)},
\label{Gloc}
\end{equation}
at each step of the standard DMFT procedure.

Eventually, we get the desired Green function in the form of (\ref{Gk}),
where $\Sigma(\varepsilon)$ and $\Sigma_{\bf p}(\varepsilon)$ are those 
appearing at the end of our iteration procedure.

To treat electron-phonon interaction for strongly correlated system we
just introduce $\Sigma_{\bf p}(\varepsilon)=\Sigma_{ph}(\varepsilon,{\bf p})$ due
to electron--phonon interaction within the usual Fr\"ohlich model.
To solve single impurity Anderson problem we use NRG\cite{NRGrev}.
All calculations are done at nearly zero temperature and at half filling.
For ``bare'' electrons we assume semielliptic DOS with half--bandwidth $D$.

\begin{figure}[t]
\begin{center}
\includegraphics[scale=0.25]{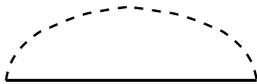}
\end{center}
\caption{ 
Migdal-like contribution to electron-phonon self-energy
included into DMFT+$\Sigma_{ph}$ scheme.
\label{migdiag}
}
\end{figure}

According to the Migdal theorem in adiabatic approximation \cite{Migdal} we can 
restrict ourselves with the simplest first order contribution to 
$\Sigma_{ph}(\varepsilon,{\bf p})$, shown by diagramm in Fig.~\ref{migdiag}.
The main advantage of this is possibility to neglect any order vertex 
corrections due electron-phonon coupling which are small over adiabatic 
parameter $\frac{\omega_D}{\varepsilon_F}\ll 1$ \cite{Migdal}.
Contribution shown in Fig.~\ref{migdiag} can be written as
\begin{eqnarray}
\Sigma_{ph}(\varepsilon,\textbf{p})=ig^2 \sum_{\omega,\textbf{k}}
\frac{\omega^2_0(\textbf{k})}{\omega^2-\omega^2_0(\textbf{k})+i\delta}\nonumber\\
\frac{1}{\varepsilon+\omega+\mu-\varepsilon({\textbf{p}+\textbf{k}})-
\Sigma(\varepsilon+\omega)-\Sigma_{ph}(\varepsilon+\omega,\textbf{p}+\textbf{k})}
\label{phse}
\end{eqnarray}
where $g$ is the usual electron-phonon interaction constant, 
$\omega_0(\textbf{k})$ is phonon dispersion, which in our case is taken 
as in the standard Debye model
\begin{equation}
\omega_0(\textbf{k})=u|\textbf{k}|,~|\textbf{k}| < \frac{\omega_D}{u}.
\label{dspec}
\end{equation}
Here $u$ is the sound velocity and $\omega_D$ is Debye frequency.

Actually $\Sigma_{ph}(\varepsilon,\textbf{p})$ defined by Eq.~(\ref{phse})
has weak momentum dependence which we can omit and continue only with
significant frequency dependence. For the Debye spectra (\ref{dspec})
Eq.~(\ref{phse}) can be rewritten as (cf. similar analysis in Ref. \cite{diagrammatics})
\begin{eqnarray}
\Sigma_{ph}(\varepsilon)=\frac{-ig^2}{4\omega_c^2} \int_{-\infty}^{+\infty}\frac{d\omega}{2\pi}
\bigl\{\omega_D^2+\omega^2ln\bigl|\frac{\omega_D^2-\omega^2}{\omega^2}\bigr|\nonumber\\
+i\pi\omega^2\theta(\omega_D^2-\omega^2)\bigr\}I(\varepsilon+\omega),
\label{phdb}
\end{eqnarray}
with
\begin{equation}
I(\epsilon)=\int_{-D}^{+D}d\xi\frac{N_0(\xi)}{E_{\varepsilon}-\xi}.
\end{equation}
where $E_{\varepsilon}=\varepsilon-\Sigma(\varepsilon)-\Sigma_{ph}(\varepsilon)$
and $\omega_c=p_Fu$ is a characteristic frequency of the order of 
$\omega_D$.
For the case of semielliptic non-interacting DOS $N_0(\varepsilon)$ with
half-bandwidth $D$ we get:
\begin{equation}
I(\epsilon)=\frac{2}{D^2}(E_{\varepsilon}-\sqrt{E_{\varepsilon}^2-D^2}),
\end{equation}
It is convenient to introduce the dimensionless electron-phonon coupling constant 
as\cite{diagrammatics}:
\begin{equation}
\lambda=g^2N_0(\varepsilon_F)\frac{\omega_D^2}{4\omega_c^2}.
\label{lambda}
\end{equation}
To simpilfy our analysis we shall not perform fully self-consistent calculations 
neglecting phonon renormalization due to EPI\cite{diagrammatics}, assuming that 
the phonon spectrum (\ref{dspec}) is fixed by the experiment.

\section{Results and discussion}
\label{results}

\begin{figure}[t]
\begin{center}
\includegraphics[angle=270,scale=0.35]{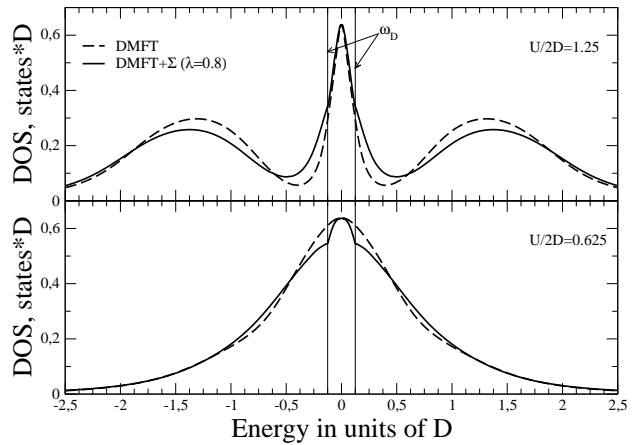}
\end{center}
\caption{ 
Comparison of DOSes obtained within standard DMFT (dashed lines) and 
DMFT+$\Sigma_{ph}$ (solid lines) methods for strong (upper panel, U/2D=1.25) 
and weak (lower panel, $U/2D=$0.625) Hubbard interaction regimes.
Dimensionless electron--phonon coupling constant $\lambda$=0.8.
\label{DOSes}
}
\end{figure}

Let us start from comparison between pure DMFT and DMFT+$\Sigma_{ph}$ 
DOSes for strong (U/2D=1.25) and weak (U/2D=0.625) Hubbard interaction
presented in Fig.~\ref{DOSes} on upper and low panels correspondingly.
Dimensionless EPI constant (\ref{lambda}) used in these calculations 
was $\lambda$=0.8 and Debye frequency  $\omega_D$=0.125D. 
In both cases we observe some spectral weight redistribution due to EPI.
For U/2D=1.25 (upper panel of Fig.~\ref{DOSes})
we see the well developed three peak structure typical for strongly correlated 
metals. In the energy interval $\pm \omega_D$ around the Fermi energy
(which is taken as zero energy at all figures below) 
there is almost no difference in the DOS quasiparticle peak line shape obtained 
from pure DMFT and DMFT+$\Sigma_{ph}$.
However outside this interval DMFT+$\Sigma_{ph}$ quasiparticle peak becomes 
significantly broader with spectral weight coming from Hubbard bands.
This broadening of DMFT+$\Sigma_{ph}$ quasiparticle peak leads
as we show below to inhibiting of metal to insulator transition.
In the case of U/2D=0.625 there are no clear Hubbard bands formed but only some 
``side wings'' are observed. 
Spectral weight redistribution on the lower panel of Fig.~\ref{DOSes} is not 
dramatic, though qualitatively different from the case of U/2D=1.25. 
Namely, main deviations between pure DMFT and DMFT+$\Sigma_{ph}$ happen in the 
interval $\pm \omega_D$, where one can observe kind of ``cap'' in 
DMFT+$\Sigma_{ph}$  DOS. Corresponding spectral weight goes to
the energies around $\pm$U, where Hubbard bands are supposed to form.
\begin{figure}[t]
\begin{center}
\includegraphics[angle=270,scale=0.35]{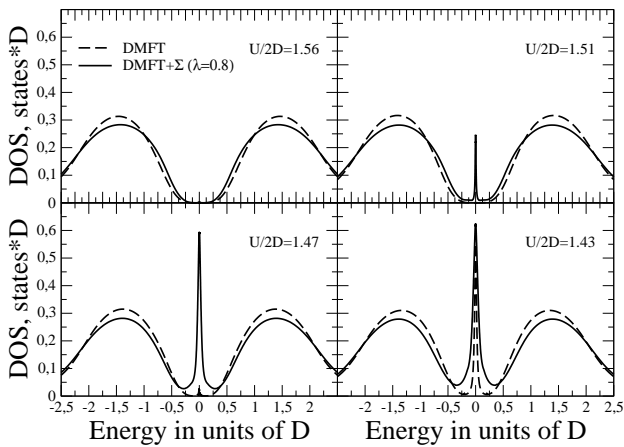}
\end{center}
\caption{ 
Sequence of DOSes obtained within standard DMFT (dashed lines) and 
DMFT+$\Sigma_{ph}$ (solid lines) methods close to metal-insulator transition 
(from top-left to bottom right) with $\lambda$=0.8.
\label{MITDOSes}}
\end{figure}

In Fig.~\ref{MITDOSes} we compare the behavior of pure DMFT and 
DMFT+$\Sigma_{ph}$ DOSes for different U/2D values close to Mott-Hubbard 
metal-insulator transition.
For U/2D=1.56 both standard DMFT and DMFT+$\Sigma_{ph}$ produce insulating 
solution. However there is some difference between these solutions. 
The DMFT+$\Sigma_{ph}$ Hubbard bands are lower and broader than DMFT ones
because of additional interaction (EPI) included.
With decrease of U for U/2D=1.51 and 1.47 we observe that DMFT+$\Sigma_{ph}$ results correspond to metallic 
state (with narrow quasiparticle peak at the Fermi level), 
while conventional DMFT still produces insulating solution.
Only around U/2D=1.43 both DMFT and DMFT+$\Sigma_{ph}$ results turn out to be 
metallic. Overall DOSes lineshape is the same as discussed above.
Thus with increase of U finite EPI slightly inhibits Mott-Hubbard transition from metallic to insulating
phase. This result is similar to what was observed for the HHM in weak EPI 
regime\cite{KolMayHew04,Jeon,Mayer04}.

\begin{figure}[t]
\begin{center}
\includegraphics[angle=270,scale=0.35]{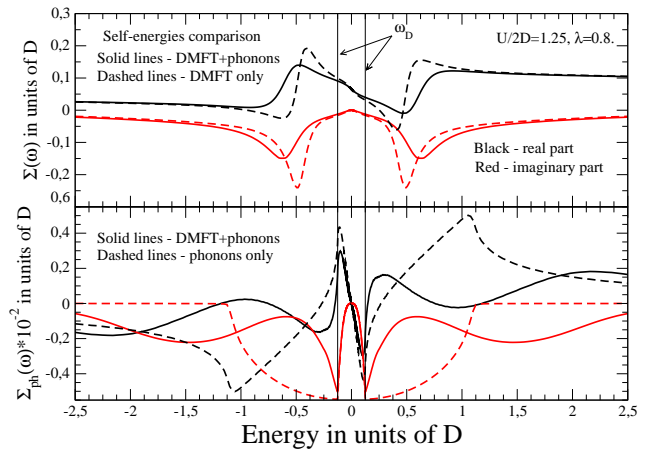}
\end{center}
\caption{(Color online) 
Upper panel --- comparison of standard DMFT self-energies $\Sigma(\varepsilon)$ 
(dashed lines) with self-energies renormalized by phonons and obtained within 
the DMFT+$\Sigma_{ph}$ approximation (solid lines).
Lower panel --- EPI self-energies $\Sigma_{ph}(\varepsilon)$.
Black lines - real parts, red lines - imaginary parts. 
$\lambda$=0.8, U/2D=1.25.
\label{sigcomp}}
\end{figure}

For more deep insight into these results let us analyze
the structure of corresponding self-energies $\Sigma(\varepsilon)$
and $\Sigma_{ph}(\varepsilon)$. In Fig.~\ref{sigcomp} we show both real and imaginary
part of these self-energies. 
EPI changes $\Sigma(\varepsilon)$ rather significantly
(see upper panel of Fig.~\ref{sigcomp}).
At the same time in $\pm \omega_D$ energy interval we find that slopes of 
real parts of both self-energies (which determines quasiparticle weight in the 
Fermi liquid theory) are almost the same, while imaginary parts are very close 
to zero. Thus quasiparticle peaks should be essentially identical in this region 
as we showed above (Fig.~\ref{DOSes}).
At energies higher than Debye frequencies Re$\Sigma(\varepsilon)$ goes steeper 
with respect to Re($\Sigma+\Sigma_{ph})$, making DMFT qusiparticle peak in DOS narrower 
above $\omega_D$ thus providing faster metal to insulator transition at $\lambda$=0.
For the case of U/2D=0.625 (not shown here) pure DMFT self-energy and those 
with the account of EPI are nearly identical. Corresponding $\Sigma_{ph}$ 
is very close to that obtained due to phonons only and shown
on lower panel of Fig.~\ref{sigcomp} with dashed lines. It produces only the 
``cap'' in the DOS around the Fermi level mentioned above. 
One can say also that such a ``cap''appears in DOS when energy interval 
2$\omega_D$ is much smaller than the quasiparticle peak width.

Now we address the issue of a sudden change of the slope of electronic 
dispersion, the so-called kinks.
It is well known that interaction of electrons with some bosonic mode
produces such a kink. In the case of EPI typical kink energy is 
just the Debye frequency $\omega_D$.
Kinks of purely electronic nature were recently reported in 
Ref.~\onlinecite{Nature}.

\begin{figure}[t]
\begin{center}
\includegraphics[scale=0.2]{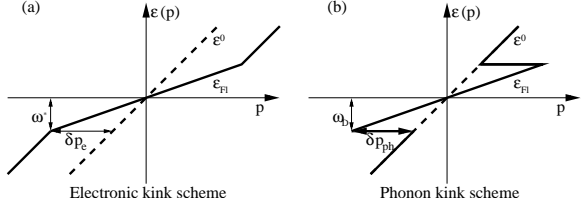}
\end{center}
\caption{
Schematic picture of pure electronic kink (panel (a)) and
phonon kink (panel (b)) in electron energy dispersion near the Fermi level.
$\varepsilon^0$ -- bare energy dispersion with no interactions
included; $\varepsilon_{Fl}$ - dispersion around the Fermi level
with electron interaction included;
$\omega^*$ - electronic kink energy; $\omega_D$ - phonon kink (Debye) energy;
$\delta p_{e}$ and $\delta p_{ph}$ -- shifts of dispersion due to pure 
electronic and phonon kinks.
\label{kinksch}}
\end{figure}

The energy of purely electronic kink as derived in Ref.~\onlinecite{Nature} 
for semielliptic bare DOS is given by
\begin{equation}
\omega^*=Z_{FL}(\sqrt{2}-1)D,
\label{wst}
\end{equation}
where D is the half of bare bandwidth and
$Z_{FL}=(1-\frac{\partial Re\Sigma)}{\partial \varepsilon}\bigr|_{\varepsilon=
\varepsilon_F})^{-1}$ is Fermi liquid quasiparticle weight.
The rough estimate of $\omega^*$ is given by the half-width of quasiparticle 
peak of DOS at its half-height.
Schematic pictures of kinks of both kinds close to the Fermi level are 
shown in Fig.~\ref{kinksch}.
Electronic kink (on the right side) is rather ``round'' and usually hard to see.
This kink is formed by the smaller slope connection of two splited branches with 
initial slope (dashed line) at energy $\pm \omega^*$. 
Far away from the Fermi level both of these branches return to the
initial dispersion.
In contrast the phonon kink produces rather sharp deviation from the initial 
dispersion at $\omega_D$, but outside $\pm \omega_D$ energy interval electron
dispersion quickly returns to the initial one.

Our calculations clearly demonstrate that electronic kinks are hardly observable
on the background of phonon kinks
(as e.g. on upper panel of Fig.~\ref{sigcomp}), and special care should be taken to separate
them by rather fine tuning of the parameters of our model. To clarify this
situation we introduce an additional characteristic of the kink --- the shift of
electron dispersion in momentum space $\delta p$ at kink energy.
From simple geometry we estimate for phonon kinks
\begin{equation}
\delta p_{ph}=\frac{\omega_D}{v_F}\lambda
\label{dpph}
\end{equation}
where $v_F$ is the bare Fermi velocity and $\lambda$ was defined in 
Eq.~(\ref{lambda}).
For electronic kink the similar estimate is
\begin{equation}
\delta p_{e}=\frac{\omega^*}{v_F^*}\bigl(1-\frac{Z_{FL}}{Z_{0}}\bigr)\equiv
\frac{\omega^*}{v_F^{*}}\lambda_e,
\label{dpe}
\end{equation}
where $Z_0$ is quasiparticle weight in the case of absence of electronic kinks
(the same as $Z_{cp}$ defined in Ref.~\onlinecite{Nature}). 
Velocity $v_F^*$ is the Fermi velocity of initial dispersion, 
but it can not be just a bare one.
As was reported in Ref.~\onlinecite{Nature} 
electronic kinks can be observed only for
rather strong Hubbard interaction when three peak structure 
in the DOS is well developed and electronic
dispersion is strongly renormalized by correlation effects.
This renormalization is determined by $\lambda_e$ defined in Eq.~(\ref{dpph}),
which can be seen as kind of dimensionless interaction constant. 
In the case when both slopes on the Fermi level and out of $\pm \omega^*$ 
energy interval are equal there will be no electronic kink at all.

\begin{figure}[t]
\begin{center}
\includegraphics[angle=270,scale=0.32]{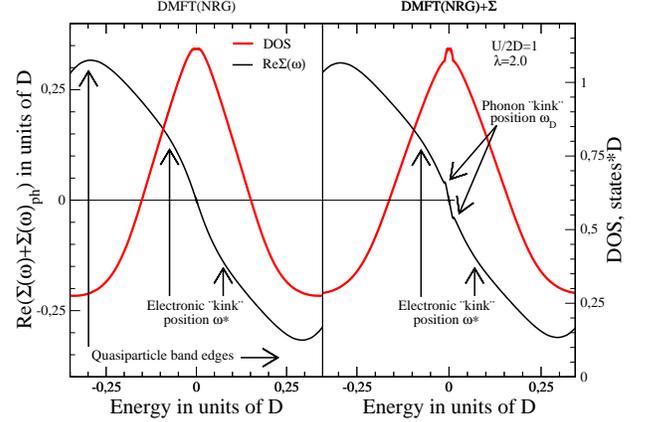}
\end{center}
\caption{(Color online)
Quasiparticle part of DOSes (see Fig.~\ref{DOSes}, upper panel) (red line) and corresponding
real part of additive self-energy Re$(\Sigma+\Sigma_{ph})$
with electron-phonon coupling switched off (left panel) and switched on 
(right panel). $\lambda$=2.0, U/2D=1.
\label{sigdos}}
\end{figure}

Now  we can choose parameters of our model
to make both kinks simultaneously visible.
First of all one should take care that $\omega_D \ll \omega^*$.
For U/2D=1 with U=3.5~eV we get $\omega^*\sim$0.1D and a reasonable
value of Debye frequency is $\omega_D\sim$0.01D.
To make phonon kink pronounced at such relatively low Debye frequency
(cf. Eq.~(\ref{dpph})) we have to increase EPI constant. 
So we take $\lambda$=2.0.
Corresponding quasiparticle peaks of the DOS together with 
Re$(\Sigma+\Sigma_{ph})$ are shown in Fig.~\ref{sigdos}: at the left panel EPI 
is switched off, while on the right panel it is switched on. 
We can see that 2$\omega^*$ is approximately
width of the quasiparticle peak of well developed three peak structure 
(see upper panel of Fig.~\ref{DOSes})
and energy position of electronic kinks are marked by arrows.
On the right side of Fig.~\ref{sigdos}, where EPI is present, 
phonon kinks at $\pm \omega_D$ are clearly
visible and well separated in energy from electronic kink position.

\begin{figure}[t]
\begin{center}
\includegraphics[angle=270,scale=0.38]{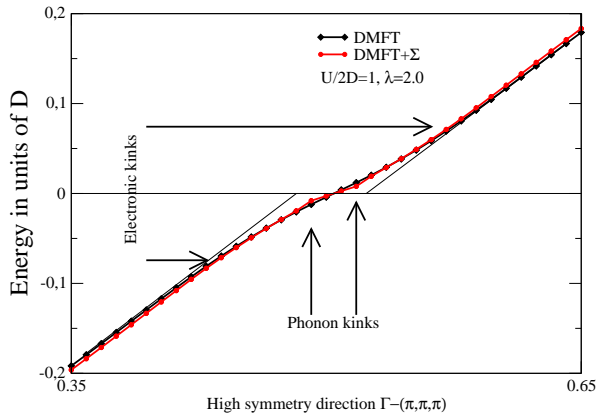}
\end{center}
\caption{(Color online)
Quasiparticle dispersions obtained from standard DMFT 
(black lines with diamonds) and DMFT+$\Sigma_{ph}$ (red lines with circles) 
around the Fermi level and along the part of high symmetry direction 
$\Gamma-(\pi,\pi,\pi)$. 
\label{ephkinks}}
\end{figure}

To demonstrate coexistence of both these types of kinks we take a look
on energy dispersion of simple cubic lattice with nearest neighbors transfers only.
Most convenient is high symmetry direction $\Gamma-(\pi,\pi,\pi)$ direction\cite{Nature}.
In Fig.~\ref{ephkinks} dispersion along this direction around Fermi level is shown.
Black line with diamonds is pure DMFT electronic spectrum, while red line with 
circles represent the result of DFMT+$\Sigma_{ph}$ calculations. 
Electronic and phonon kinks are marked with arrows.

\begin{figure}[b]
\begin{center}
\includegraphics[scale=0.42]{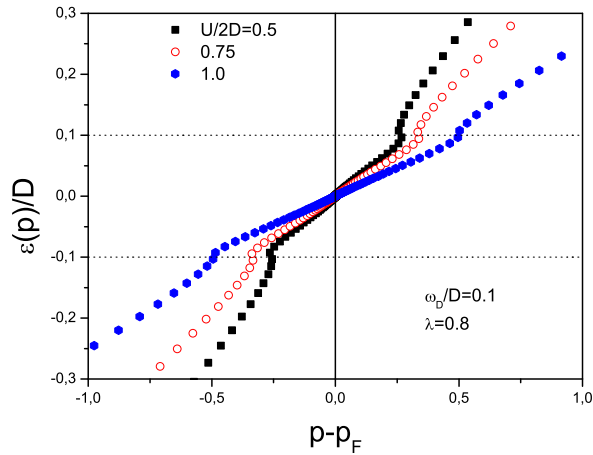}
\end{center}
\caption{(Color online)
Quasiparticle dispersions around Fermi level
with phonon kinks obtained from DMFT+$\Sigma_{ph}$ calculations
for different interaction strengths U/2D=\ 0.5,\ 0.75,\ 1.0;
\ $\lambda=0.8$, $\omega_D$=0.1D. 
\label{Ukinks}}
\end{figure}

Finally we address to the behavior of phonon kinks in electronic spectrum
as function of Hubbard interaction U. As U/2D ratio grows Fermi velocity in 
Eq.~(\ref{dpph}) goes down, so that momentum shift of kink position $\delta p$ 
moves away from $p_F$, while kink energy remains at $\omega_D$. This is
confirmed by our direct DMFT+$\Sigma_{ph}$ calculations producing the
overall picture of spectrum evolution shown in Fig.~\ref{Ukinks}. 

\section{Conclusion}
\label{conclusion}

This work is a first attempt to analyze strongly correlated electrons, treated
within DMFT approach to the Hubbard model, interacting with Debye phonons.
EPI is treated within the simplest (Migdal theorem) approach in adiabatic
approximation, allowing the neglect of vertex corrections. DMFT+$\Sigma_{ph}$
approach allows us to use the standard momentum space representation for
phonon self-energy (\ref{phse}), while the general structure of DMFT equations
remains intact.

Mild EPI leads to rather insignificant changes of electron density of states,
both in correlated metal and in Mott--insulator state, slightly inhibiting
metal to insulator transition with increase of U.

However, kinks in the electronic dispersion due to EPI dominate for the most
typical values of the model parameters, making kinks of purely electronic
nature, predicted in Ref. \cite{Nature}, hardly observable. Special care (fine
tuning) of model parameters is needed to separate these anomalies in
electronic dispersion in strongly correlated systems.

We have also studied phonon kinks evolution with the strength of electronic
correlations demonstrating the significant drop in the slope of electronic 
dispersion close to the Fermi level with the growth of Hubbard interaction $U$.

We believe that these results may be of importance in further studies of the
evolution of electronic spectra in highly corretaed systems, such as e.g
copper -- oxides.

\section{Acknowledgments} 

We are grateful to Th. Pruschke for providing us with his effective NRG code.

This work is partly supported by RFBR grants 08-02-00021, 08-02-91200 and 
08-02-00712. It was performed within the framework of the programs of 
fundamental research of the Russian Academy of Sciences (RAS) ``Quantum physics 
of condensed matter'' and of the Physics Division of RAS  ``Strongly correlated 
electrons in solid states''. 
I.N. acknowledges Russian Science Support Foundation and Grant of President of
Russia for young PhD.

\end {document}